\documentstyle[epsfig,11pt]{article}

\def\beq{\begin{equation}}
\def\eeq{\end{equation}}
\def\pd{\partial}
\def\Phid{\Phi^\dagger}
\title{\bf BFKL pomeron in the external field of the nucleus
in (2+1)-dimensional QCD}
\author{M.A.Braun, A.N.Tarasov  \\
St.Petersburg State University, Russia}
\begin{document}
\maketitle

{\bf Abstract}
The behaviour of the pomeron propagator in the external nuclear field is
studied in the (2+1)-dimensional QCD. It is shown that in the
physically interesting case
when the field does not vanish at large rapidities the propagator in the
field vanishes much faster than in the vacuum, in agreement with the results
found in the local Regge-Gribov model. However if the nuclear field
vanishes at high rapidities the field does not change the behaviour of the
pomeron propagator.

\section{Introduction}
In the framework of the perturbative QCD at small $x$ and large $N_c$
the strong interaction is realized by the exchange of hard BFKL pomerons
which interact via the 3-pomeron vertex governing their
splitting and fusion. One can sum all the corresponding fan diagrams in the
quasi-classical approximation by means of the Balitski-Kovchegov (BK)
equation for $\gamma$ (h)A scattering \cite{bal,kov} or by means of
a pair of equations for AB scattering introduced in ~\cite{braun1}.
With the growth of energy the role of pomeron loops becomes important and
one has to search for methods to take them into account. In our previous
publications we noted that calculations of loops may become easier if
one starts with the perturbative approach inside the nucleus from the start.
At least in the local Regge-Gribov model  the nuclear field transforms
the supercritical pomeron with the intercept greater than unity
into a subcritical one with the intercept smaller than unity
~\cite{BT1}. As a result with the growth of energy
the loop becomes relatively smaller, so that it can be
calculated perturbatively. 

This result can be easily understood from the
structure of the model in the nuclear surrounding. The Lagrangian
of the Regge-Gribov model has a form
\beq
{\cal L}=\Phid(\frac{\pd}{\pd y}+H)\Phi+V\Phid\Phi(\Phid+\Phi)
\eeq
where $\Phi$ and $\Phid$ are quantum fields which describe the pomeron,
 $H$ is the free "Hamiltonian" and $V$ - the 3-pomeron vertex.
In the quasi-classical approximation one has $\Phid=0$ and $\Phi=\Phi_0$,
the latter given by the sum of all fan diagrams corresponding to the
interaction with the nucleus. To calculate quantum correction one can
shift the field $\Phi\to \Phi+\Phi_0$ and thus study the model in the
nuclear background field $\Phi_0$. This introduces a new interaction
$V\Phi_0{\Phid}^2$ corresponding to the annihilation of a pair of pomerons
and, most important, the pomeron propagator must now be calculated in the
background field $\Phi_0$. It turns out that this propagator, which in
the vacuum exponentially grows with rapidity, in the background nuclear field
exponentially vanishes. As a result in the background
field pomeron loops do not grow with rapidity and can be easily controlled.

Of course the problem is to see if this result can be generalized to the
perturbative QCD pomeron.
Our numerical calculations have shown that the QCD pomeron propagator
in the external field of the nucleus also vanishes at high rapidities
in contrast to its behaviour in the vacuum \cite{BT2}. Unfortunately the realistic
QCD in 3+1 dimensions does not allow to confirm this result analytically,
since both the relevant BK equation and the BFKL
equation in the external field do not allow for analytic solutions.
As was shown a some time ago the situation is improved in the
(2+1)-dimensional QCD, where a solution to the BK equation can be
obtained in a form which analytically shows its rapidity dependence
~\cite{BFL}.

In this note we use this known solution to study the behaviour of the
BFKL propagator in the external field given by this solution and
corresponding to the field inside a big nucleus.
Our results show that
this behaviour, as well as the behaviour of the nuclear field itself,
very much depend on the initial conditions for evolution. If they are chosen
in such a way that at large rapidities the fan-diagram-field tends to unity
(the dipole $S$-matrix vanishes) then solutions of the BFKL equation in this
field also vanish in this limit. However for different initial conditions,
for which  the fan diagram-field vanishes at large rapidities (the dipole
$S$-matrix tends to unity), this field does not
change the behaviour of  the solutions of the BFKL equation.
A striking result of our study is that the interval of the coordinates
in which the  BFKL propagator in the external field is different from zero
diminishes with the fall of energy. As a consequence pomerons cannot
form loops in the nuclear field. This implies that the quasi-classical
approach gives the complete solution to the
quantum field theory of interacting pomerons in the nuclear field in 2+1
dimensions.

\section{Main equations}

As found in ~\cite{BFL} the kernel of the BK equation (the rescaled
triple pomeron vertex) is changed for the (2+1)-dimensional space
according to
\beq
\frac{\alpha_sN_c}{2\pi^2}\frac{r_{21}^2}{r_{23}^2r_{31}^2}
\to 2\alpha_sN_c\theta(r^{max}_{2,1}-r_3)\theta(r_3-r^{min}_{2,1}),
\label{vertex}
\eeq
where $r^{max(min)}_{2,1}$ is the maximal (minimal) of the
coordinates $r_2$ and $r_1$; $r_{21}=r_2-r_1$ etc. This vertex is a constant
different from zero only when $r_3$ lies between $r_2$ and $r_1$.
This allows to write the BK equation in a simple form.
Let $S_{r_2r_1}(y)$ be the $S$-matrix for the interaction of a dipole
stretched between spatial (1-dimensional) points $r_2>r_1$ at rapidity
$Y$. The BK equation for $S$, which sums all the fan diagrams, is
\beq
\frac{\pd S_{r_2r_1}}{\pd y}=
\int_{r_1}^{r_2}dr_0\Big(S_{r_2r_0}S_{r_0r_1}-
S_{r_2r_1}\Big),\ \ y=2\pi \bar{\alpha}Y,
\label{eqs}
\eeq
where $Y$ is the rapidity.
This equation can be further simplified if we introduce a function
\beq
 \Psi_{r_2r_1}(y)=e^{r_{21}y}S_{r_2r_1},\
\ r_{21}=r_2-r_1\geq 0.
\eeq
The equation for $\Psi$ is
\beq
\frac{\pd \Psi_{r_2r_1}}{\pd y}=
\int_{r_1}^{r_2}dr_0\Psi_{r_2r_0}\Psi_{r_0r_1}.
\label{eqg}
\eeq
Considering $\Psi_{r_2r_1}(y)$ as a triangular matrix
\beq
\Psi_{r_2r_1}(y)=<r_2|\Psi(y)|r_1>,\ \ r_2\geq r_1,
\eeq
we can rewrite Eq. (\ref{eqg}) as
\beq
\frac{\pd \Psi(y)}{\pd y}=\Psi^2(y),
\label{eqg1}
\eeq
with a solution satisfying the initial condition
\beq
\Psi_{r_2r_1}(y)\Bigr|_{y=0}=S_{r_2r_1}(y)\Bigr|_{y=0}=\Psi_{r_2r_1}(0)
\eeq
and given by the matrix formula
\beq
\Psi(y)=\Psi(0)[1-y\Psi(0)]^{-1}.
\eeq

To obtain the equation for the BFKL pomeron  inside the nucleus we
first rewrite Eq. (\ref{eqs}) in terms of the scattering amplitude
$\Phi_{r_2r_1}(y)$ defined by
\beq
S_{r_2r_1}(y)=1-\Phi_{r_2r_1}(y).
\eeq
We get
\beq
\frac{\pd \Phi_{r_2r_1}}{\pd y}=
\int_{r_1}^{r_2}dr_0\Big(
\Phi_{r_2r_0}+\Phi_{r_0r_1}-\Phi_{r_2r_1}-
\Phi_{r_2r_0}\Phi_{r_0r_1}\Big).
\label{eqphi}
\eeq
This scattering amplitude represents the sum of fan diagrams connecting
the projectile with the nucleus target. This is precisely the  field
created by the nuclear background.

Dropping the non-linear term one obtains the BFKL equation in 2+1
dimensions in the vacuum
\beq
\frac{\pd P_{r_2r_1}}{\pd y}=
\int_{r_1}^{r_2}dr_0\Big(
P_{r_2r_0}+P_{r_0r_1}-P_{r_2r_1}\Big).
\label{eqn0}
\eeq
The BFKL equation in the nuclear field is obtained when one
adds to the right-hand side terms which correspond to the interaction
of the pomeron with the nuclear field via the triple pomeron vertex
\beq
\frac{\pd P_{r_2r_1}}{\pd y}=
\int_{r_1}^{r_2}dr_0\Big(
P_{r_2r_0}+P_{r_0r_1}-P_{r_2r_1}-
\Phi_{r_2r_0}P_{r_0r_1}-P_{r_2r_0}\Phi_{r_0r_1}\Big).
\label{eqn}
\eeq
Expressing $\Phi$ via $S$ we rewrite this equation as
\beq
\frac{\pd P_{r_2r_1}}{\pd y}=
\int_{r_1}^{r_2}dr_0
\Big(
S_{r_2r_0}P_{r_0r_1}+P_{r_2r_0}S_{r_0r_1}-
P_{r_2r_1}\Big).
\label{eqn1}
\eeq

Finally we introduce
\beq
Q_{r_2r_1}(y)=e^{r_{21}y}P_{r_2r_1}(y),
\eeq
for which the equation simplifies to
\beq
\frac{\pd Q_{r_2r_1}}{\pd y}=
\int_{r_1}^{r_2}dr_0
\Big(
\Psi_{r_2r_0}Q_{r_0r_1}+Q_{r_2r_0}\Psi_{r_0r_1}\Big)
\label{eqm}
\eeq
or in the matrix notation
\beq
\frac{\pd Q(y)}{\pd y}=\Big\{Q(y),\Psi(y)\Big\}.
\label{eqm1}
\eeq

Solution to this equation with the initial condition
\[Q(y,y')\Big|_{y=y'}=Q(y',y')\]
is found to be
\[
Q(y,y')=\Psi^{-1}(y')\Psi(y)Q(y',y')\Psi(y)\Psi^{-1}(y')\]
\beq\equiv T(y,y')Q(y'y')T(y,y'),
\label{solq}
\eeq
where we define triangular matrix
\beq
T(y,y')=\Psi^{-1}(y')\Psi(y)=\Psi(y)\Psi^{-1}(y').
\label{tdef}
\eeq
Note that $\Psi(y)$ and $\Psi^{-1}(y')$ commute at any $y$ and $y'$,
since they both are functions of the same matrix $\Psi(0)$.

In the following we shall consider a simple case when the nuclear 1-dimensional density is constant ("nuclear matter"). Then the initial
BK function $\Psi_{r_2r_1}(0)$ will depend only on the distance between
the gluons
\beq
\Psi_{r_2r_1}(0)=\Psi_{r_{21}}(0).
\eeq
As a result $\Psi_{r_2r_1}(y)$ will depend only on the distance
$\rho_{21}$ at all values of $y$ and consequently $T_{r_2r_1}(y,y')$
will also depend only on the distance $r_{21}$. As a matrix
\beq
T_{r_2r_1}(y,y')=\theta(r_{21})T_{r_{21}}(y,y').
\eeq

\section{The Green function of the BFKL equation (\ref{eqn1})}
We shall first study the Green function $G_{r_2r_1|r'_2r'_1}(y,y')$
which satisfies
 \beq
 G_{r_2r_1|r'_2r'_1}(y',y')=
 \delta(r_2-r'_2)\delta(r_1-r'_1)
 \label{inigg}
\eeq
and we assume that $r_2=\max\{r_2,r_1\}$ and $r'_2=\max\{r'_2,r'_1\}$.
Note that this is not the pomeron propagator, which will be studied later.
To find $G$ we have
to search for a solution of the equation in variables $r_2$ and $r_1$
for $P$ which satisfies (\ref{inigg}).
In terms of
 \[
F_{r_2r_1}(y,y')=
e^{y r_{21}}G_{r_2r_1|r'_2r'_1}(y,y')\]
we have to find a solution of    Eq. (\ref{eqn1}) which satisfies
 \[F_{r_2r_1}(y',y')=e^{y'r'_{21}}
 \theta(r_{21})\delta(r_2-r'_2)\delta(r_1-r'_1).\]
Obviously we should also have $r'_2\geq r'_1$.

According to (\ref{solq}) it is given by
\[
F_{r_2r_1}(y,y')=e^{y'r'_{21}}
\theta(r_{21})\theta(r'_{21})
\int dr_0dr'_0
T_{r_2r_0}(y,y')\delta(r_0-r'_2)\delta(r'_0-r'_1)
T_{r'_0r_1}(y,y')\]\beq=
e^{y'r'_{21}}\theta(r_{21})
\theta(r'_{21})T_{r_2r'_2}(y,y')T_{r'_1r_1}(y,y').
\label{solf}
\eeq
The Green function itself will be given by
\beq
G_{r_2r_1|r'_2r'_1}(y,y')
=e^{y'r'_{21}-yr_{21}}\theta(r_{21})\theta(r'_{21})
T_{r_2r'_2}(y,y')T_{r'_1r_1}(y,y').
\label{solg}
\eeq
This expression is different from zero only in the interval
\beq G_{r_2r_1|r'_2r'_1}(y,y')\neq 0
 \ \ {\rm only\ if}\ \ r_1<r'_1<r'_2<r_2.
\label{interval}
\eeq
Here $r_2$ and $r'_2$ are the maximal coordinates of the initial and
final gluons and $r'_1$ and $r'$ are their minimal coordinates.
From (\ref{interval}) one concludes that the  interval of coordinates
where $G$ is different from zero is restricted on both sides.
This interval at $y$ should be wholly inside the one at the initial rapidity $y'$. This remarkable property will have far-reaching
consequences for the whole model, as will be discussed in the following.

For the considered case of nuclear matter when
$
T_{r_2r_1}=T_{r_{21}}
$
the Green function $G$ can be presented in the
form which explicitly shows its symmetry in the initial and final gluons
\beq
G_{r_2r_1|r'_2r'_1}(y,y')=e^{y's'-ys} T_{s_2}(y,y')T_{s_1}(y,y').
\label{solgg}
\eeq
where
\[
s=\max\{r_2,r_1\}-\min\{r_2,r_1\}\ \
s'=\max\{r'_2,r'_1\}-\min\{r'_2,r'_1\},\]\beq
s_2=\max\{r_2,r_1\}-\max\{r'_2,r'_1\},\ \
s_1=\min\{r'_2,r'_1\}-\min\{r_2,r_1\}
\label{llp}
\eeq
and $s$, $s'$, $s_1$ and $s_2$ are all to be non-negative
so that $s_2+s_1=s-s'\geq 0$.

At $y'=0$ we have
\beq
T_{r_2r_1}(y,0)=
\Big([1-y\Psi(0)]^{-1}\Big)_{r_2r_1}.
\label{defx}
\eeq
From the equation
\beq
\int_{r_1}^{r_2}dr_0T_{r_2r_0}(y,0)\Big(\delta(r_0-r_1)-
y\Psi_{r_0r_1}(0)\Big)=\delta(r_2-r_1)
\eeq
we conclude that
\beq
T_{r_2r_1}(y,0)=\delta(r_2-r_1)+Y_{r_2r_1}(y),
\eeq
where $Y(y)$ is a smooth function of $r_2$ and $r_1$.
Presence of the $\delta$-function allows to get the correct right-hand side,
which otherwise would be zero at $r_2=r_1$. Using
 the fact that the BK solution depends only on the difference
$r_{21}$ we find
\beq
T_{r_{21}}(y)=\delta(r_{21})+Y_{r_{21}}(y).
\eeq
To find $Y_r(y)$ we set up an evolution equation in $y$.
Obviously
\beq
\frac{\partial T_r(y,0)}{\partial y}=
\Big(\Psi(0)T^2(y,0)\Big)_r=
\int_0^r dr_1\Psi_{r-r_1}(0)\int_0^{r_1} dr_2T_{r_1-r_2}(y,0)
T_{r_2}(y,0),
\eeq
or in terms of $Y$
\beq
\frac{\partial Y_r(y)}{\partial y}=\Psi_{r}(0)+
2\int_0^{r} dr_1\Psi_{r-r_1}(0)Y_{r_1}(y)+
\int_0^r dr_1\Psi_{r-r_1}(0)\int_0^{r_1} dr_2Y_{r_1-r_2}(y)Y_{r_2}(y)
\label{evol}
\eeq
with the initial condition
\beq
Y_r(0)=0.
\label{ini1}
\eeq
From this evolution equation one can numerically find $Y_r(y)$ at any
$y$ by the standard Runge-Kutta procedure.
%%%%%%%%%%%%%%%%%%%%%%%%%%%%%%%%%%%%%%%%%%55555
At arbitrary $y$ and $y'$
\beq
T(y,y')=[1-y'\Psi(0)][1-y\Psi(0)]^{-1}.
\eeq
From this we find
\beq
T_{r}(y,y')=\delta(r)+X_{r}(y,y'),
\eeq
where $X(y,y')$ is a smooth function linear in $y'$:
\beq
X_r(y,y')=Y_r(y)-y'\Big[\Psi_r(0)+
\int_0^{r} dr' \Psi_{r-r'}Y_{r'}(y)\Big].
\eeq

In absence of the interaction with the nucleus $S$-matrix
corresponding to the BK equation turns into unity. In terms
of the triangular matrices it implies
\beq
S^{(0)}_{r_2r_1}(y)=\theta(r_2-r_1)
\eeq
and does not depend on $y$.  From this we find in the vacuum
\beq
T^{(0)}_r(y,y')=\delta(r)+(y-y')e^{yr}\theta(r).
\label{t0}
\eeq
Using this  and
(\ref{solg}) we obtain the Green function $G^{(0)}$ in the vacuum:
\[
G^{(0)}_{r_2r_1|r'_2r'_1}(y,y')=
e^{-(y-y')r'_{21}}\Big\{
\delta(r_{22'})\delta(r_{1'1})\]\beq+
(y-y')\delta(r_{1'1})\theta(r_{22'})+
(y-y')\delta(r_{22'})\theta(r_{1'1})+
(y-y')^2\theta(r_{22'})\theta(r_{1'1})\Big\},
\eeq
where it is assumed that $r_{22'},r_{1'1}\geq 0$.
This expression coincides with the one previously obtained
in ~\cite{BFL}. As expected in absence of the external field the
Green function depends only on the difference $y-y'$.
%%%%%%%%%%%%%%%%%%%%%%%%%%%%%%%%%%%%%%%%%%%%%%%

\section{Pomeron propagator in the nuclear field}
In contrast with  the Green function
$G$ defined by the equation
\beq
\Big(\frac{\pd}{\pd y}+H\Big)G=1,
\eeq
where $H$ is the BFKL Hamiltonian in the external field,
the pomeron propagator $g$ is defined by
\beq
\Big(\frac{\pd}{\pd y}+H\Big)g=p_1^{-2}p_2^{-2},
\eeq
so that it is related to the Green function considered previously
as
\beq
g(r_2,r_1|r'_2,r'_1)=\nabla_1^{-2}\nabla_2^{-2}G(r_2,r_1|r'_2,r'_1)=
G(r_2,r_1|r'_2,r'_1)\nabla_{1'}^{-2}\nabla_{2'}^{-2},
\label{gg}
\eeq
or equivalently
\beq
\nabla_1^2\nabla_2^2g(r_2,r_1|r'_2,r'_1)=
g(r_2,r_1|r'_2,r'_1){\nabla'_1}^2{\nabla'_2}^2=
G(r_2,r_1|r'_2,r'_1),
\label{gg1}
\eeq
where in the last equality the derivatives act on the left.
Obviously relations (\ref{gg1}) determine $g$ non-uniquely. One
can add to it any function linear in $r_2-r'_2$ or in $r_1-r'_1$.
Our choice will consist in selecting $g$ to be different from zero
in the same intervals of initial and final gluon coordinates as $G$,
namely $s,\ s',\ s_1,\ s_2\geq 0$ where $s$ are defined by (\ref{llp}).
As we shall find the propagator $g$ defined in this way at $y>y'$
vanishes at large separation of points. Linear terms that can be added
to it and different from zero outside the interval (\ref{interval})
violate this condition and seem to us unacceptable on formal grounds,
since correspond to strong infrared singularities at all values of $y$.

To make our choice we notice that
in the one-dimensional transverse coordinate space function
$h(r)=\nabla^{-2}\delta(r)$ can be presented as
\beq
h(r)=r\theta(r).
\eeq
Indeed we  have
\[\nabla^2h(r)=h''(r)=
(r\theta(r))''=r\delta'(r)+2\delta(r)=\delta(r).\]

This implies that the Green function $g(y,r_2,r_1|y',r'_2,r'_1)$
may be chosen to satisfy the BFKL equation
\beq
\Big(\frac{\pd}{\pd y}+H\Big)g(y,r_2,r_1|y',r'_2,r'_1)=
(r_2-r'_2)\theta(r_2-r'_2)(r'_1-r_1)\theta(r'_1-r_1),
\label{eqgg}
\eeq
where it is assumed $r_{21},\ r'_{21}\geq 0$.
Note the inverse order of the initial and final coordinates in
the second factor on the right. It is this order that guarantees
the desired property of $g$ to be different from zero only in the
interval (\ref{interval}).

We can solve this equation in the same manner as when searching for
the Green function $G$. We introduce  matrix
\beq
f_{r_2r_1}(y,y')=\theta(r_{21})e^{yr_{21}}
g_{r_2r_1|r'_2r'_1}(y,y')
\eeq
with  $r_{21},\ r'_{21}\geq 0$.
It will satisfy Eq. (\ref{eqgg})
with the initial condition
\beq
f_{r_2r_1}(y',y')=\theta(r_{21})
e^{y'r_{21}}\theta(r'_{21})
(r_2-r'_2)\theta(r_2-r'_2)
(r'_1-r_1)\theta(r'_1-r_1).
\eeq
According to Eq. (\ref{solq})
the solution is given by
\beq
f_{r_2r_1}(y,y')=
\sum_{r_0,r'_0}\theta(r_{00'})T_{r_2r_0}(y,y')
r_{02'}\theta(r_{02'})
r'_{10}\theta(r'_{10})e^{y'r_{00'}}
T_{r'_0r_1}(y,y'),
\eeq
where we use the notation $r_{00'}=r_0-r'_0$ and $r'_{10}=r'_1-r'_0$.

Taking into account that $T_{r_2r_1}$ depends only on the distance
$r_{21}$ between the points we find
\[
g_{r_2r_1|r'_2r'_1}(y,y')=
\theta(r_{21})\theta(r'_{21})e^{-r_{21}(y-y')}
\int dr\int dr'\theta(r_{21}-r-r')\theta(r_{22'}-r)
\theta(r_{1'1}-r')\]\beq
e^{-y'(r+r')}(r_{22'}-r)(r_{1'1}-r')T_r(y,y')T_{r'}(y,y').
\eeq
Integration over $r$ and $r'$ is restricted by the last two
$\theta$-functions to the interval
\[ 0<r<r_{22'},\ \ O<r'<r_{1'1}.\]
From this we conclude
\[r+r'<r_{22'}+r_{1'1}=r_{21}-r'_{21}<r_{21} \]
so that $\theta(r_{21}-r-r')=1$. As a result this expression factorizes
\beq
g_{r_2r_1|r'_2r'_1}(y,y')=
e^{-r_{21}(y-y')}U_{r_{22'}}(y,y')U_{r_{1'1}}(y,y'),
\label{solgp}
\eeq
where we have defined a matrix
\beq
U_{r}(y,y')=
\int_0^{r}dr' (r-r')T_{r'}(y,y')e^{-y'r'}.
\eeq
In the notation symmetric in the initial and final gluons this
can be rewritten as
\beq
g_{r_2r_1|r'_2r'_1}(y,y')=e^{-s(y-y')} U_{s_1}(y,y')U_{s_2}(y,y').
\label{solgg1}
\eeq
where $s$, $s'$, $s_1$ and $s_2$ are defined by Eqs. (\ref{llp}).

We have seen that
\beq
T(y,y')=[1-y'\Psi(0)][1-y\Psi(0)]^{-1}
\eeq
and that the second factor contains a $\delta$-contribution
\beq
[1-y\Psi(0)]^{-1}=1+Y(y).
\eeq
From this we find
\beq
T(y,y')=1+X(y,y'),\ \ X(y,y')=Y(y)-y'\Psi(0)-y'\Psi(0)Y(y),
\eeq
so that $T_{r_2r_1}$ contains a $\delta$ term.
If $T_{r_2r_1}$ depends only on the distance $r_{21}$
we find
\beq
T_{r}(y,y')=\delta(r)+X_{r}(y,y'),
\eeq
where $X_r(y,y')$ is a smooth function of $r$ linear in $y'$:
\beq
X_r(y,y')=Y_r(y)-y'\Big[\Psi_r(0)+
\int_0^{r} dr' \Psi_{r-r'}Y_{r'}(y)\Big].
\eeq

Separating the $\delta$ function from $T$ one gets
\beq
U_{r}(y,y')=r+
\int_0^{r}dr' X_{r'}(y,y')(r-r')e^{-y'r'}.
\eeq

In absence of the nuclear field, using (\ref{t0}) one obtains
\beq
U^{(0)}_r(y-y')=\frac{e^{(y-y')r}-1}{y-y'}.
\label{u0}
\eeq
From this using (\ref{solgp}) we find the vacuum propagator,
which naturally depends only on the difference $y-y'$, which we denote as
simply $y$
\beq
g_{r_2r_1|r'_2r'_1}(y)=\frac{1}{y}\Big\{
e^{-yr_{21}}+e^{-yr'_{21}}-e^{-yr_{2'1}}-e^{-yr_{21'}}\Big\},
\eeq
where it is assumed that $r_1\leq r'_1\leq r'_2\leq r_2$.

\subsection{Numerical results}
Both the nuclear field  characterized by the BK $S$-matrix
$S_{r_2r_1}(y)$ and the pomeron propagator are determined by
the coupling to the nucleus contained in the BK $S$-matrix
at $y=0$. In the nuclear matter on physical grounds at $r=0$ one requires
$S_{r}(0)_{r=0}=1$ corresponding to vanishing of integration for
zero-dimensional dipoles. As noted in ~\cite{BFL} the behaviour of $S$
with the growth of energy crucially depends on the behaviour of $S_r(0)$
$S_r(0)_{r\to 0}\sim 1-cr^\gamma$. In particular for $\gamma<1$ it was
found that the asymptotical behavior at $y>>1$ was
\beq
S_{r}(y)\sim e^{-c\Gamma(1+\gamma)y^{1-\gamma}r}.
\label{asym1}
\eeq
On the other hand for $\gamma>1$ the $S$-matrix was expected to tend to
unity  in the large region $r<<y^{\gamma-1}$.

\begin{figure}
\hspace*{2 cm}
\epsfig{file=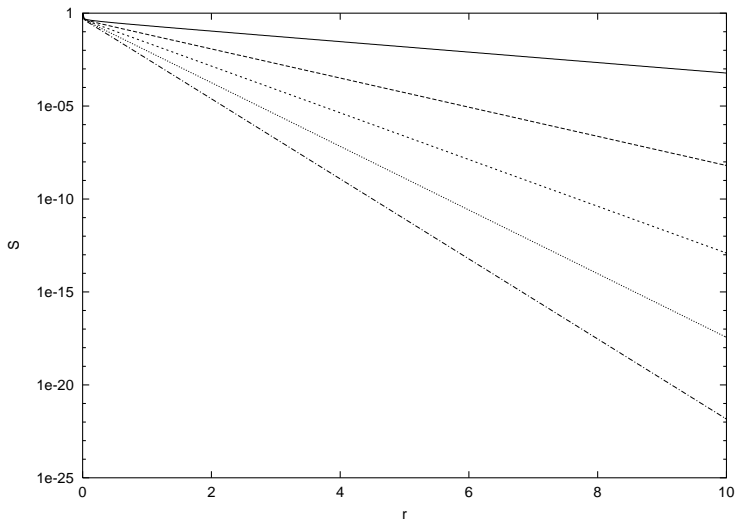,width=10 cm}
\caption{$S$-matrix $S_r(y)$ evolved by the BK equation from the initial
function Eq. (\ref{ini}) with $\gamma=0.1$ as a function of $r$.
Curves from top to bottom correspond to $y=1,$ 3, 5, 7 and 9}
\label{fig1}
\end{figure}

\begin{figure}
\hspace*{2 cm}
\epsfig{file=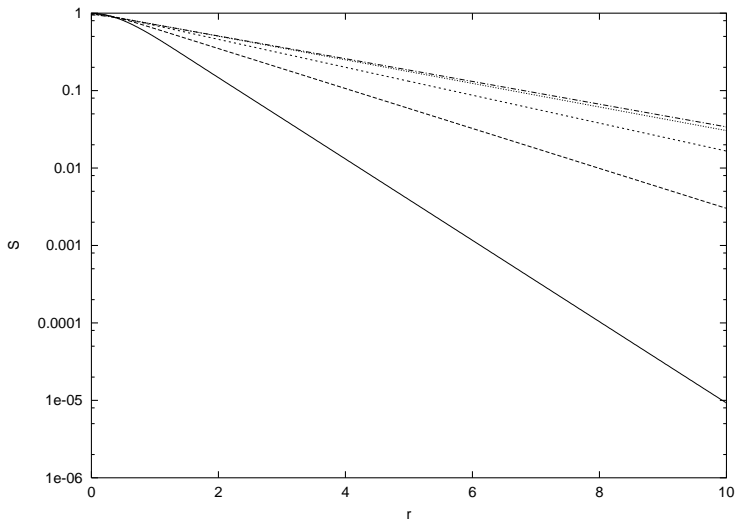,width=10 cm}
\caption{Same as Fig. \ref{fig1} with $\gamma=2.0.$
Curves from bottom to top correspond to $y=1,$ 3, 5, 7 and 9}
\label{fig2}
\end{figure}

In our calculations we have taken $S_r(0)$ in the form
\beq
S_r(0)=e^{-r^\gamma}
\eeq
Numerical results show that for $\gamma<1$ the asymptotical behaviour
(\ref{asym1}) is well fulfilled. In Fig. \ref{fig1} we show the
behavour $S_r(y)$ with $\gamma=0.1$ for values of rescaled rapidity
$y=1,3,5,7$ and 9. In fact they are well described by (\ref{asym1})
practically at all values of $r$ and $y$.
However for $\gamma>1$ we find the asymptotical behaviour somewhat different
from what was expected in ~\cite{BFL}. In Fig. \ref{fig2} we present
$S_r(y)$ with $\gamma=2$ at the same rapidities $y=1,3,5,7$ and 9.
They all fall onto the exponential dependence on $r$
\beq
S_r(r)\sim e^{-b(y)r}.
\label{ini}
\eeq
In contrast to the case $\gamma<1$ the slope $b(y)$ falls with $y$ but
not to zero but to the finite value $\sim 0.33$. The same picture
is observed with still higher values of $\gamma$.

Passing to the pomeron propagator we note that it depends on three
independent variables.
Assuming that $r_1\leq r'_1\leq r'_2\leq r_2$,
they can be taken as any three of the four differences
$r_{21}$, $r'_{21}$, $r_{22'}$ and $r_{1'1}$, taking non-negative values and
constrained by the relation
\beq
r'_{21}+r_{22'}+r_{1'1}=r_{21}.
\label{variables}
\eeq

To see both $r-$ and $y-$ dependence
we shall first study
integrals of $g$ at $y'=0$ over $r_{22'}$ and $r_{1'1}$ with fixed
$r_{21}=r$
\beq
g_1(y,r)=e^{-yr}\int_0^r dr_1U_{r_1}(y,0)\int_0^{r-r_1} dr_2
U_{r_2}(y,0)
\eeq
and with fixed $r'_{21}=r$
\beq
g_2(y,r))=e^{-yr}\int_0^{r_m-r}dr_1U_{r_1}(y,0)e^{-yr_1}
\int_0^{r_m-r-r_1}dr_2U_{r_2}(y,0)e^{-yr_2},
\eeq
where $r_m$ is the upper limit of the integration, (formally infinite,
but finite in the course of calculations and taken $r_m=20$).

\begin{figure}
\hspace*{2 cm}
\epsfig{file=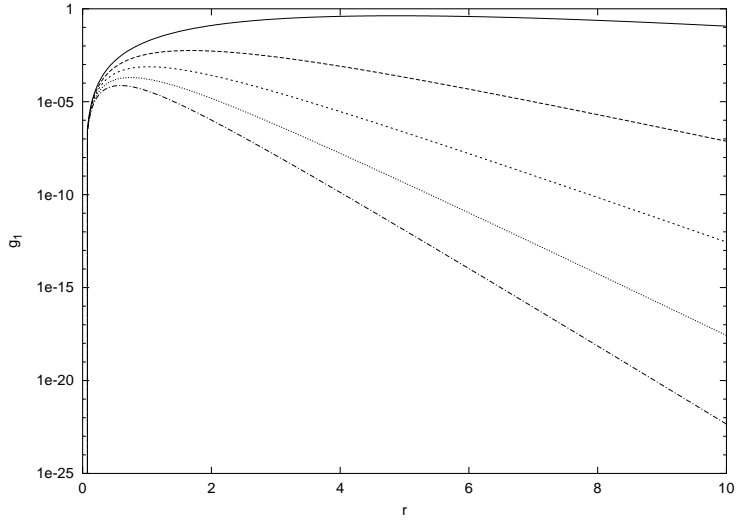,width=10 cm}
\caption{Pomeron propagator integrated over the final gluon coordinates
$g_1(y,r)$ as a function of $r$.
The initial BK function Eq. (\ref{ini}) is taken with $\gamma=0.1$.
Curves from top to bottom correspond to $y=1,$ 3, 5, 7 and 9}
\label{fig3}
\end{figure}

\begin{figure}
\hspace*{2 cm}
\epsfig{file=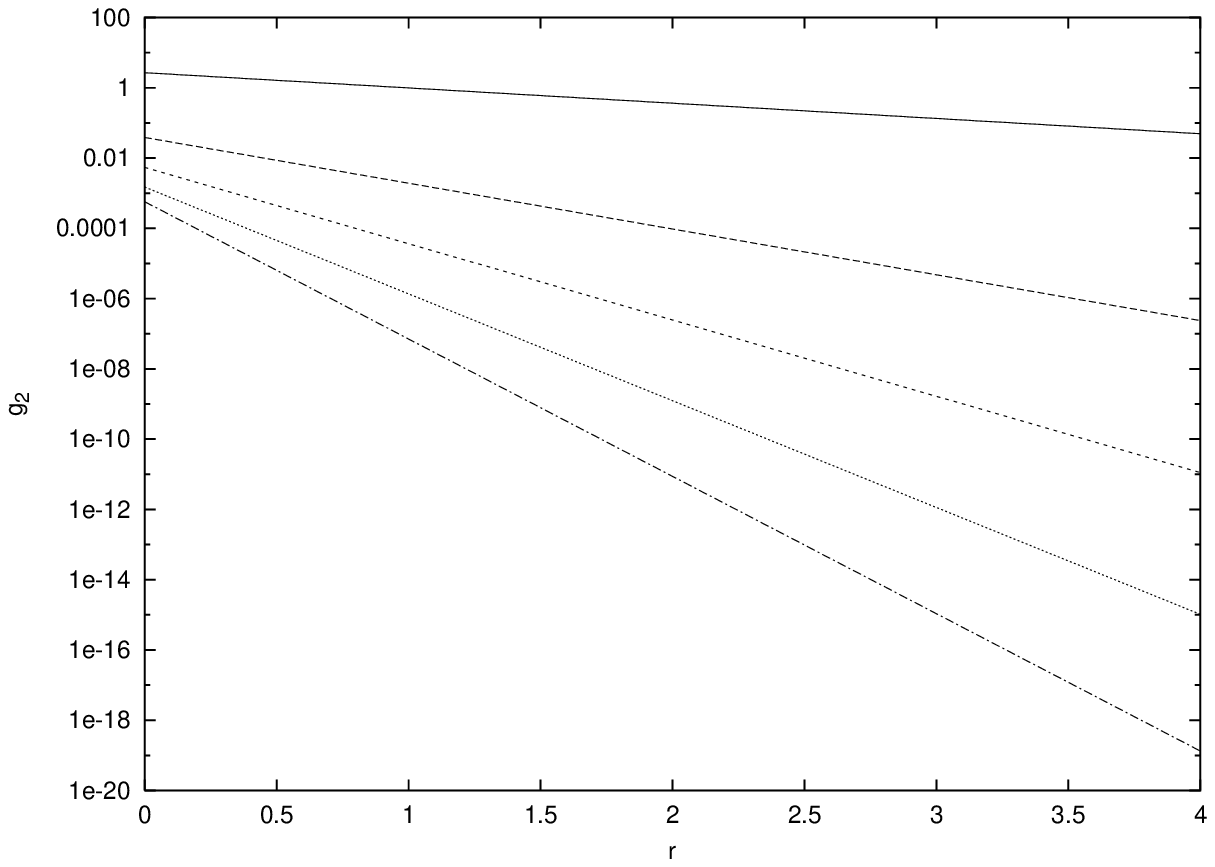,width=10 cm}
\caption{Pomeron propagator integrated over the initial gluon coordinates
$g_2(y,r)$ as a function of $r$.
The initial BK function Eq. (\ref{ini}) is taken with $\gamma=0.1$.
Curves from top to bottom correspond to $y=1,$ 3, 5, 7 and 9}
\label{fig4}
\end{figure}

\begin{figure}
\hspace*{2 cm}
\epsfig{file=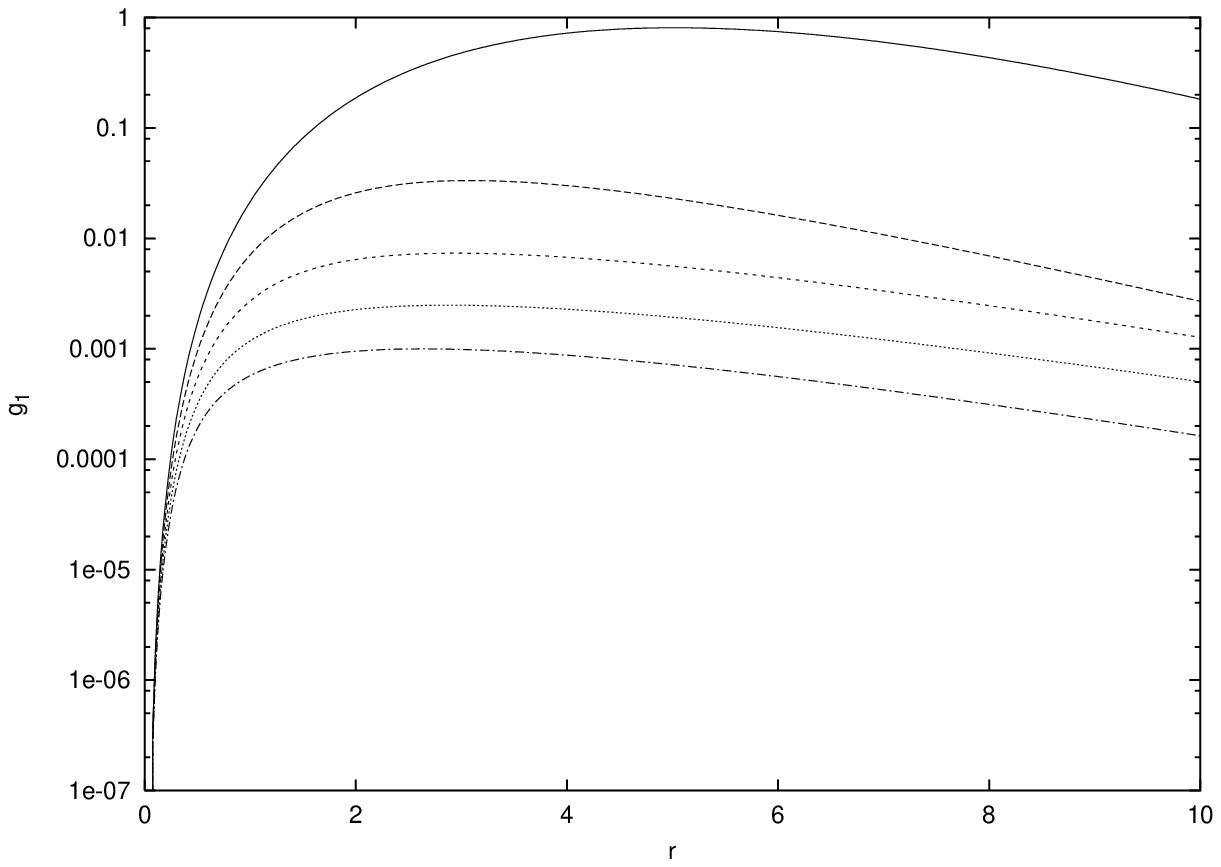,width=10 cm}
\caption{Same as Fig. \ref{fig3} with $\gamma=2.0$ in
the initial BK function Eq. (\ref{ini})
Curves from top to bottom correspond to $y=1,$ 3, 5, 7 and 9}
\label{fig5}
\end{figure}

\begin{figure}
\hspace*{2 cm}
\epsfig{file=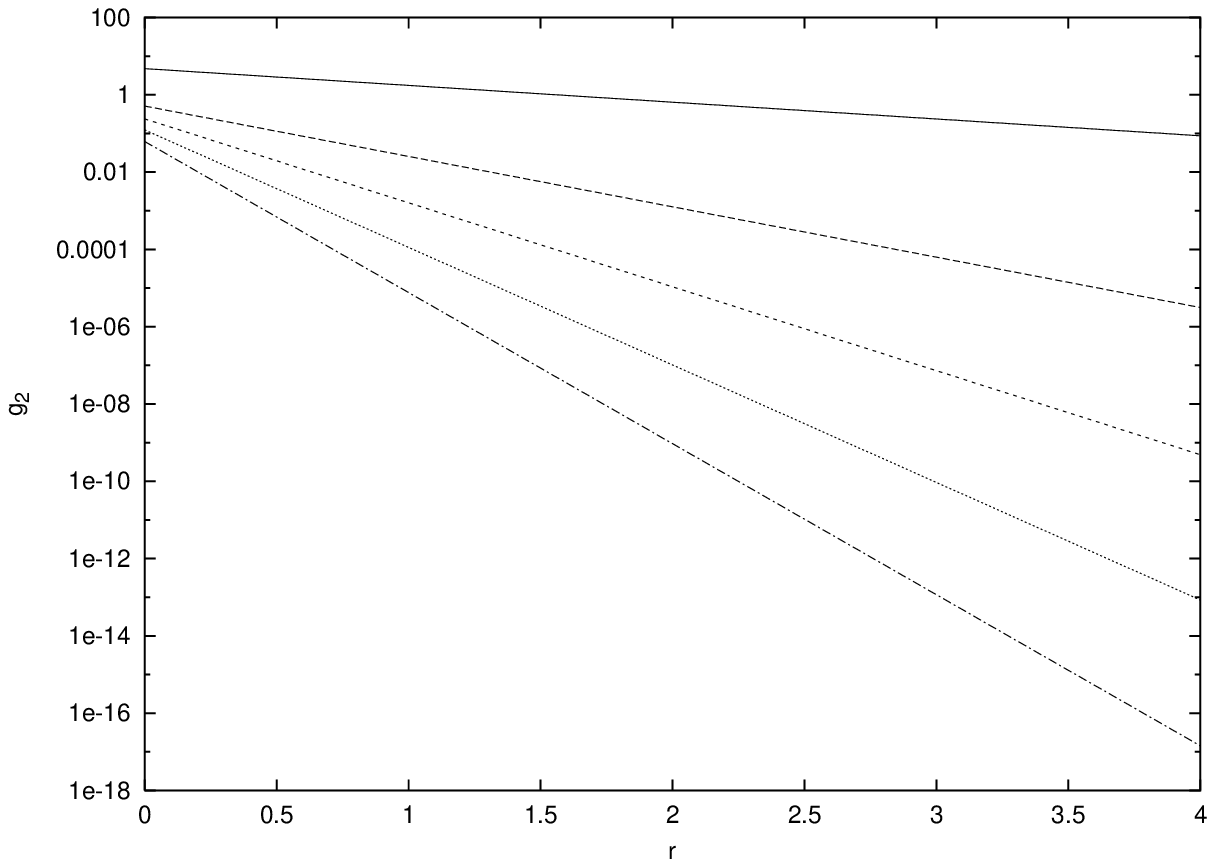,width=10 cm}
\caption{Same as Fig. \ref{fig4} with $\gamma=2.0$ in
the initial BK function Eq. (\ref{ini})
Curves from top to bottom correspond to $y=1,$ 3, 5, 7 and 9}
\label{fig6}
\end{figure}

\begin{figure}
\hspace*{2 cm}
\epsfig{file=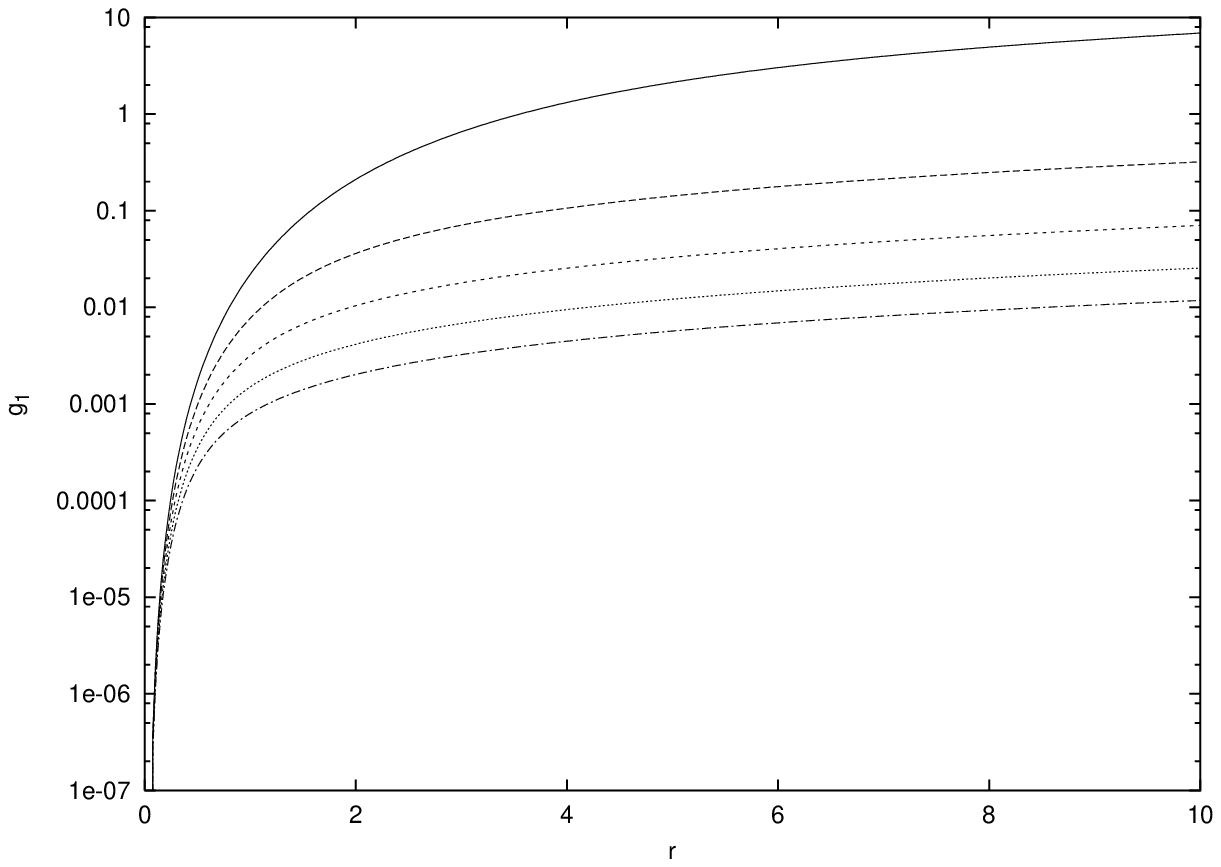,width=10 cm}
\caption{Same as Fig. \ref{fig3} for the vacuum case.
Curves from top to bottom correspond to $y=1,$ 3, 5, 7 and 9}
\label{fig7}
\end{figure}

\begin{figure}
\hspace*{2 cm}
\epsfig{file=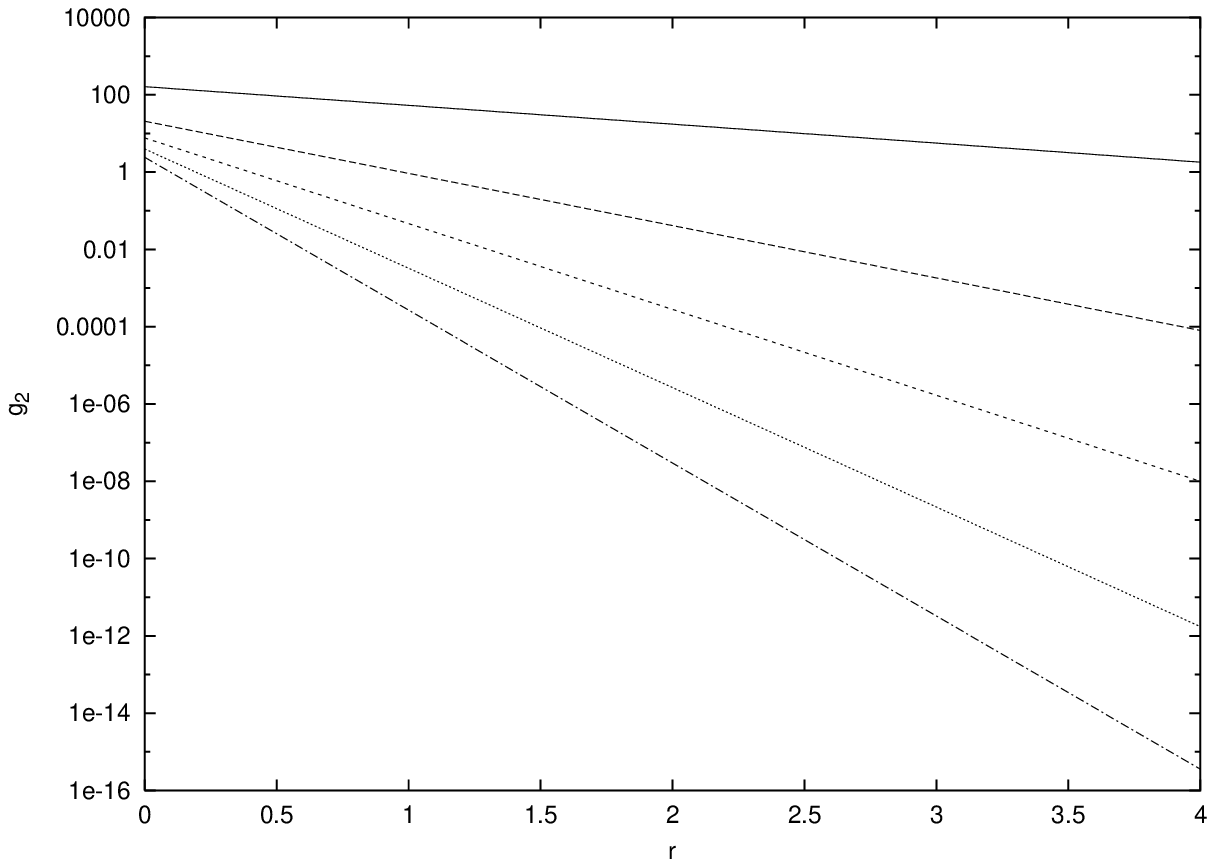,width=10 cm}
\caption{Same as Fig. \ref{fig4} for the vacuum case.
Curves from top to bottom correspond to $y=1,$ 3, 5, 7 and 9}
\label{fig8}
\end{figure}

Our numerical results for $g^{(1)}_r$ and $g^{(2)}_r$ with $\gamma=0.1$
and $y=1,3,5,7$ and 9
are shown in Figs. \ref{fig3} and \ref{fig4} respectively. Similar
results with $\gamma=2.0$ are illustrated in Figs. \ref{fig5} and
\ref{fig6}.
To compare in Fig. \ref{fig7} and \ref{fig8} we plot the same quantities
for the vacuum case.
In all cases the integrated propagator
diminishes with energy and coordinate.

\begin{figure}
\hspace*{2 cm}
\epsfig{file=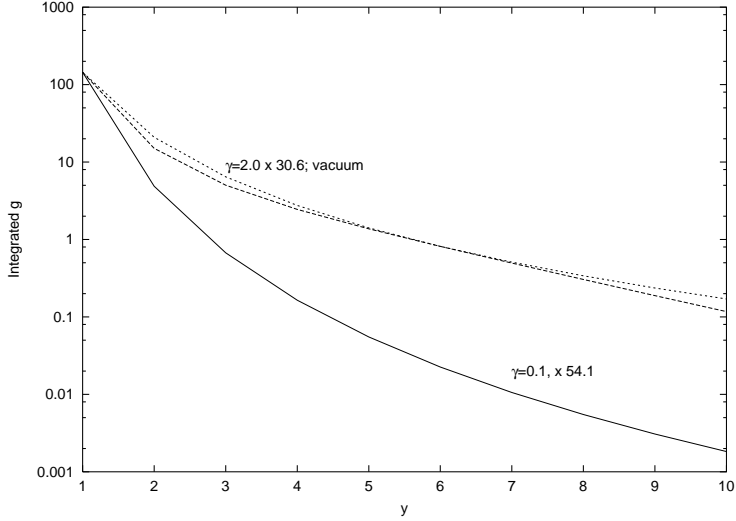,width=10 cm}
\caption{Fully integrated propagator $g(y,0)$.
The two curves with $\gamma=0.1$ (the lower one) and $\gamma=2.0$
(one of the two practically coinciding upper curves) are rescaled
to have the same value with the vacuum curve (the other of the
two upper curves) at $y=1$}
\label{fig9}
\end{figure}

To clearly see the influence of the field and the role of value of $\gamma$
in Fig. \ref{fig9} we show values of $g(y_,0)$ integrated over all three
independent variables from (\ref{variables}). To facilitate the comparison
results with $\gamma=0.1$ and $\gamma=2.0$ were rescaled to coincide
with the vacuum result at $y=1$.
We observe a clear distinction between the cases $\gamma<1$ and $\gamma>1$
previously stressed in ~\cite{BFL}. With $\gamma<1$ we   find that the
propagator in the nuclear field indeed goes to zero much faster
than in the vacuum as advocated in our previous studies in the 3+1
dimensional case. In contrast with $\gamma>1$ we do not observe any
influence of the field. This goes in line with the assertion in
~\cite{BFL} that this case corresponds to asymptotic vanishing of the
dipole interaction.

\section{Loops}
In 3+1 dimensions in the theory of interacting BFKL pomerons the
interaction is given by
\beq
S_I=\frac{2\alpha_s^2N_c}{\pi}
\int dy\frac{d^2r_1d^2r_2d^2r_3r_{12}^2}{r_{23}^2r_{31}^2}
[\nabla_1^2\nabla_2^2\phi^+(y,r_1,r_2)]
\cdot \phi(y,r_2,r_3)\phi(y,r_3,r_1)
+\Big(h.c\Big).
\eeq
Accordingly  the pomeron self-mass $\Sigma$ is given by
\beq
\Sigma(y,r_2,r_1|y',r'_2,r'_1)=-\Big(\frac{2\alpha_s^2N_c}{\pi}\Big)^2
\int\frac{d^2r_3 r_{12}^2}{r_{23}^2r_{31}^2}
\frac{d^2r'_3 r_{1'2'}^2}{r_{2'3'}^2r_{3'1'}^2}
g(y,r_2,r_3|y',r'_2,r'_3)g(y,r_1,r_3|y',r'_1,r'_3)
\label{sigma}
\eeq
(the minus sign reflects the absorptive character of the triple
pomeron interaction).

Passing to 2+1 dimensions we substitute the vertex function
$r_{12}^2/r_{23}^2r_{13}^2$ by the difference of $\theta$ functions,
as found in \cite{BFL} and indicated in (\ref{vertex}). The
pomeron self-mass becomes
\[
\Sigma(y,r_2,r_1|y',r'_2,r'_1)=-(8\pi N_c\alpha_s^2)^2
\int_{min\{r_2,r_1\}}^{\max\{r_2,r_1\}}dr_3
\int_{min\{r'_2,r'_1\}}^{\max\{r'_2,r'_1\}}dr'_3\]\beq
g(y,r_2,r_3|y',r'_2,r'_3)g(y,r_3,r_1|y',r'_3,r'_1)
\label{sigmaa}
\eeq
It is trivial to  see that this expression is zero
due to the properties of the pomeron propagator.
In fact take $r_2>r_1$ and $r'_2>r_1'$. The first propagator
is different from zero only at $r'_3>r_3$.
and the  second propagator is different from zero only at
$r_3>r'_3$. So their product is zero.

\begin{figure}
\hspace*{2 cm}
\epsfig{file=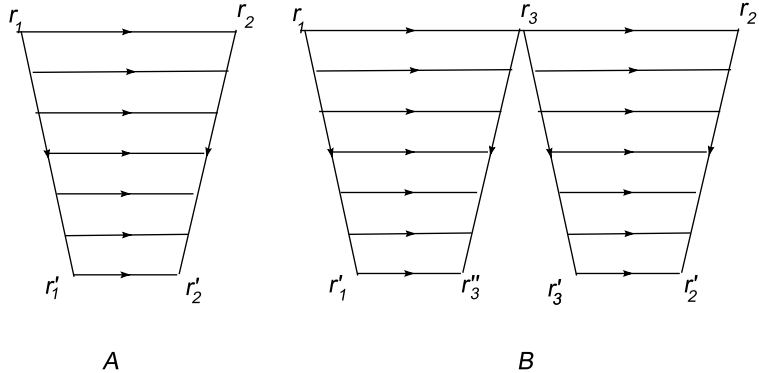,width=10 cm}
\caption{Schematic regions in the $(y,r)$ plane
where the pomeron propagator (a) or
a pair of propagators after splitting (b) are different from zero}
\label{fig10}
\end{figure}

This circumstance is obvious since the interval where the propagator
is different from zero is extending on both sides with the growth of $y$
as schematically shown in Fig. \ref{fig10},$a$. Splitting the pomeron in
two generates the picture shown in Fig. \ref{fig10}.$b$.
It is evident that after splitting as the rapidity diminishes the
two new pomerons can never interact with each other (points $r'_3$
and $r''_3$ in Fig. \ref{fig10},$b$ cannot coincide).
So loops do not exist in our theory.

\section{Conclusions}

In 2+1 dimensions solutions of both the BK equation describing the
nuclear field and BFKL equation in this field can be obtained in the
closed form, expressed via infinite-dimensional matrices on the
points dividing the coordinate interval. This makes it possible to find
both the Green function of the BFKL equation and the pomeron propagator.
It is important that the initial condition for the propagator at $y=y'$
in principle admits addition of terms linear in one of the two
final coordinates. We have proposed a choice which in our opinion is
physically motivated, since it has no infrared singularities at $y-y'>0$.
With this choice calculation of the propagator shows that the influence
of the nuclear field very much depends on the behaviour of this field.
If at large rapidities the nuclear $S$-matrix tends to zero (black disk
limit) the field indeed makes the propagator fall with rapidity
much faster than in the vacuum. On the other hand if the nuclear $S$-matrix
tends to unity at high energies (zero interaction with the nucleus) the
field in fact does not change the behaviour of the propagator in comparison
to the nuclear case.

An important consequence of our choice of the propagator is that the region
where it is different from zero does not change with energy and restricted
to the interval of final coordinates which contains inside the one of the
initial coordinates. As a result  pomerons cannot form loops of any sort.
Thus with this choice the quasi-classical solution of the BK equation gives
the complete solution of the dipole interaction with the nucleus.
The question if other choices of the propagator are admissible from
some points of view remains open and will be considered in future
studies.

\end{document}